\documentclass{article}
\usepackage{graphicx} 
\usepackage[colorlinks=true, linkcolor=blue, citecolor=blue, urlcolor=blue]{hyperref}
\usepackage{xcolor}
\usepackage{amsmath}
\usepackage{latexsym}
\usepackage{caption} 
\usepackage[a4paper, margin=1.7cm]{geometry}
\usepackage{subcaption}
\usepackage{tikz}
\usepackage{tikz-3dplot}
\usetikzlibrary{decorations.pathmorphing, arrows.meta}
\usepackage{subcaption}  
\usepackage{float}
\usepackage{authblk}

\newcommand{\vbf}{\mathbf{v}}
\newcommand{\avg}[1]{\left\langle #1\right\rangle}
\newcommand{\mDP}{m_{A'}}
\newcommand{\dd}{\mathrm{d}}
\newcommand{\vesc}{v_\mathrm{esc}}
\newcommand{\erf}{\mathrm{erf}}
\newcommand{\fR}{f_{\rm R}}

\newcommand{\dpc}{\chi \ m_{A}^{2}}
\newcommand{\wdp}{\omega_{DP}}
\newcommand{\cov}{C_{\text{overlap}}}
\newcommand{\intk}{A'(|k|, \theta_{k}, \phi_{k})}
\newcommand{\bessel}{J_{l'}(|k| r)}
\newcommand{\TM}{\mathrm{TM}}
\newcommand{\TE}{\mathrm{TE}}

\title{Polarization and Spectral Response of Dark Photon Haloscopes}
\author[1,2]{Miriam Bulliri\thanks{\href{mailto:mbulliri@unisa.it}{mbulliri@unisa.it}}}
\author[1,2]{Luca Visinelli\thanks{\href{mailto:lvisinelli@unisa.it}{lvisinelli@unisa.it}}}

\affil[1]{Dipartimento di Fisica ``E.R.\ Caianiello'', Universit\`a degli Studi di Salerno, Via Giovanni Paolo II 132, 84084 Fisciano (SA), Italy}
\affil[2]{Istituto Nazionale di Fisica Nucleare, Gruppo Collegato di Salerno, Sezione di Napoli, Via Giovanni Paolo II 132, 84084 Fisciano (SA), Italy}

\date{\today}

\begin{document}
\maketitle
\begin{abstract}
Dark photon dark matter (DPDM) can excite both transverse magnetic and transverse electric modes of a resonant electromagnetic (EM) cavity even in the absence of an external magnetic field. We derive the spectral response of a vacuum cavity to a nonrelativistic DPDM field with a general momentum distribution and polarization density matrix, accounting for the motion and orientation of the detector. We obtain analytic overlap factors for the cylindrical $\TM_{0n0}$ and $\TE_{1np}$ families, including an exact finite momentum expression for the former, and evaluate their resonant frequencies and wall loss geometrical factors for ongoing and planned cavity experiments. The fundamental axial and transverse modes have comparable maximum form factors but different resonant frequencies. Finite momentum effects provide a distinct source of directional dependence, particularly for modes whose homogeneous field overlap vanishes.
\end{abstract}

\section{Introduction}
\label{sec:introduction}

Light bosonic fields constitute compelling dark matter (DM) candidates, as their cosmological abundance can originate from nonthermal production mechanisms and their large occupation numbers allow for a classical field description. Among these candidates, massive vector bosons associated with an additional Abelian gauge symmetry or dark photons (DPs) can interact with the Standard Model through kinetic mixing with the electromagnetic (EM) field~\cite{Holdom:1985ag, Nelson:2011sf, Arias:2012az, Caputo:2021eaa}. Several mechanisms can account for the cosmological abundance of vector DM, including misalignment~\cite{Nelson:2011sf, Arias:2012az, Alonso-Alvarez:2019ixv}, the amplification of quantum fluctuations during inflation~\cite{Graham:2015rva}, parametric resonance~\cite{Dror:2018pdh, Co:2018lka}, and radiation from cosmic string networks~\cite{Long:2018nsl}. These mechanisms predict different polarization properties, which can be further affected by gravitational evolution during structure formation~\cite{Amin:2022pzv,Chen:2024vgh,Amaral:2024tjg,Chen:2026dst}. The polarization of the local DP population can provide information on both its cosmological origin and subsequent evolution.

Resonant EM cavities or haloscopes were originally proposed to detect Galactic axions through their conversion into photons in an external magnetic field~\cite{Sikivie:1983ip}. Kinetic mixing allows DPs to excite EM cavity modes even in the absence of an external magnetic field, enabling the reinterpretation of axion searches in terms of vector DM~\cite{Arias:2012az,Gelmini:2020kcu,Ghosh:2021ard,Caputo:2021eaa, McAllister:2022ibe, Chang:2025ahb}. Dedicated DP searches have been conducted with superconducting radio frequency cavities~\cite{Nguyen:2019xuh, Cervantes:2022epl, Yao:2024fie}. Other cavity haloscopes, including QUAX, provide further opportunities for vector DM searches~\cite{QUAX:2024fut}. Unlike axions, DPs possess an intrinsic polarization that determines their coupling to the electric field of the cavity modes. A preferred polarization direction in Galactic coordinates induces a characteristic sidereal modulation of the signal~\cite{Caputo:2021eaa,Amaral:2024tjg}. Moreover, TM and TE modes can provide complementary sensitivity to different polarization components, motivating the use of orthogonal and degenerate EM modes for DP detection and polarization reconstruction~\cite{Navarro-Madrid:2025qtm}.

The finite velocity dispersion of Galactic DM implies that the field possesses a finite coherence length and coherence time, motivating a description in terms of stochastic Fourier amplitudes and phases~\cite{Gelmini:2020kcu, Amaral:2024tjg,Yao:2024fie}. The resulting spectral distribution depends on the local DM phase space distribution, which is modified by the motion of the detector and solar gravitational focusing~\cite{Alenazi:2006wu, Kim:2021yyo}. Directional effects associated with the finite momentum of the DM field have been investigated for axion haloscopes~\cite{Knirck:2018knd}, while the intrinsic polarization of vector DM introduces additional directional degrees of freedom. Resonant cavities have also been considered as telescopes sensitive to relativistic DPs originating from the decay or annihilation of other dark sector particles~\cite{SHANHE:2024tpr}. Here, we focus instead on nonrelativistic DPs constituting the Galactic DM halo.

In this work, we derive the spectral and directional response of resonant EM cavities to a stochastic DP field with a general momentum distribution and polarization density matrix, retaining the spatial variation of the field across the cavity and the finite linewidth of the EM resonances. We identify the conditions under which the spectral dependence associated with the local velocity distribution and the geometrical dependence associated with polarization factorize. We derive analytical expressions for the form factors of cylindrical cavity modes and investigate the complementary polarization response of TM and TE modes, with particular relevance to large volume resonators such as those developed for the FINUDA magnet for Light Axion SearcH (FLASH) experiment at the Frascati National Laboratories~\cite{Alesini:2023qed}.

The paper is organized as follows. In Sec.~\ref{sec:polarization}, we discuss the origin and gravitational evolution of vector DM polarization. In Sec.~\ref{sec:formfactor}, we introduce the stochastic description of the local DP field and its polarization density matrix. In Sec.~\ref{sec:haloscope}, we derive the EM response and spectral power of a resonant cavity, subsequently applying the formalism to cylindrical resonators. We adopt natural units $\hbar=c=1$.

\section{A case of interest: Dark Photon dark matter from misalignment}\label{sec:polarization}

In the scalar misalignment scenario, the DM field $\phi$ is governed by the Lagrangian
\begin{equation}
    \mathcal{L}_{i} = \frac{1}{2} \partial_{\mu}\phi \partial^{\mu}\phi - \frac{1}{2} m_{\phi}^{2} \phi^{2} + \mathcal{L}_{I}\,,
\end{equation}
where $\mathcal{L}_{I}$ describes possible additional interactions. When $H\gg m_\phi$, the field is approximately frozen at its initial value, with negligible velocity. The onset of oscillations occurs approximately when $3 H(t_{1}) = m_{\phi}(t_{1}) \equiv m_{1}$. Once the oscillations become adiabatic, the evolution of the field can be described using the WKB approximation,
\begin{equation}
    \phi(t) \sim \phi_{1}(\frac{m_{1}a_{1}^{3}}{m_{\phi}(t)a^{3}(t)})^{1/2} \cos\left(\int_{t_{1}}^{t} m_{\phi}(t') \dd t' + \varphi_1\right)\,,
\end{equation}
where $a_1=a(t_1)$ and $\phi_1$ and $\varphi_1$ specify the amplitude and phase of the oscillations. The same misalignment mechanism for a vector field $X^{\mu}$ can lead to different results because its equation of motion differs from the scalar case~\cite{Arias:2012az, Nelson:2011sf}. A nonminimal interaction with gravity can be introduced through the dimensionless coupling $\kappa$,
\begin{equation}
    \label{eq:lagrangian}
    \mathcal{L} = - \frac{1}{4}X_{\mu \nu} X^{\mu \nu} + \frac{m^{2}_{\gamma'}}{2} X_{\mu} X^{\mu} + \frac{\kappa}{12} R X_{\mu} X^{\mu}\,,
\end{equation}
where $X_{\mu\nu}=\partial_\mu X_\nu-\partial_\nu X_\mu$ is the field strength tensor and $R$ is the Ricci scalar. For the homogeneous physical spatial field $\mathcal X_i=X_i/a$ in an expanding universe, the corresponding equation is
\begin{equation}
    \ddot{\mathcal X}_{i} + 3 H \dot{\mathcal X}_{i} + \left[\mDP^2+(1-\kappa)(\dot H+2H^2)\right]\mathcal X_{i} = 0\,.
\end{equation}
The case $\kappa=1$ gives a scalar-like background equation, although the stability of perturbations in theories with nonminimal curvature couplings requires a separate investigation. The evolution of the condensate can also be affected by resonant conversion in the primordial plasma. Here, we do not assume a specific cosmological production mechanism.

The polarization of vector DM can be affected by the subsequent evolution of the field during structure formation~\cite{Amin:2022pzv, Chen:2024vgh, Amaral:2024tjg, Chen:2026dst}. In the nonrelativistic limit, the spatial components of a real Proca field can be expressed in terms of a slowly varying complex field $\boldsymbol\Psi$,
\begin{equation}
    A'_{i} = \frac{1}{\sqrt{2\mDP}}\left(\Psi_{i} e^{-i\mDP t} + \Psi^*_{i}e^{i\mDP t}\right)\,.
\end{equation}
At leading order in the nonrelativistic expansion and in the presence of a weak gravitational potential $\Phi_N$, the evolution of $\boldsymbol\Psi$ is governed by the Schr\"odinger--Poisson equations,
\begin{equation}
    \label{eq:sp}
    \begin{split}
    &i \partial_t \boldsymbol \Psi = - \frac{\nabla^{2}}{2\mDP} \boldsymbol \Psi+ \mDP \Phi_N \boldsymbol \Psi\,, \\
    &\nabla^{2} \Phi_N = 4 \pi G \mDP (\boldsymbol \Psi^{\dagger} \boldsymbol\Psi-n)\,,
    \end{split}
\end{equation}
where $n$ is the mean number density given an appropriate homogeneous background subtraction. The spin in this limit is
\begin{equation}
    \mathbf S = i \int \dd^{3} x\, \boldsymbol\Psi \times \boldsymbol\Psi^{*}\,.
\end{equation}
Here, a nonzero spin density corresponds to elliptical polarization in general, not necessarily circular polarization. The appearance of a polarized soliton in simulations does not imply that the entire Galactic halo has a universally aligned circular polarization. 
For a virialized DP population, the Compton period $T_{C}= 2\pi / \mDP$ sets the rapid carrier oscillations, while the coherence time is of order $1/(\mDP\sigma_v^2)$ in terms of the characteristic velocity dispersion of the DP population $\sigma_v$. For the mass range relevant to resonant microwave cavities, both timescales are much shorter than the sidereal period. This hierarchy allows us to average over the rapid oscillations and stochastic fluctuations of the field while retaining the slow temporal variation of the cavity orientation with respect to the Galactic polarization. Motivated by the possibility that the local DP population retains a preferred polarization direction, in the following we describe its polarization state through a general density matrix, without imposing a specific cosmological production mechanism.

\section{Stochastic description of the Galactic dark photon field}
\label{sec:formfactor}

We first define the reference frames in our analysis. The Galactic axes point toward the Galactic center, along the direction of Galactic rotation, and toward the north Galactic pole, respectively. In the ecliptic frame, the $x$ axis points toward the vernal equinox, the $z$ axis toward the north ecliptic pole, and the $y$ axis completes the right-handed coordinate system. Similarly, the equatorial frame is defined by the direction of the vernal equinox and the north celestial pole, which identify the $x$ and $z$ axes, respectively. Finally, we introduce the cavity frame, in which the $z$ axis is parallel to the cylinder axis and the $x$ and $y$ axes span the transverse plane. We denote by $\mathcal R_{\rm gal\to cav}(t)$ the rotation matrix that transforms vector components from Galactic to cavity coordinates.

Given its large occupation number, DPDM can be described as a classical field consisting of a superposition of plane waves with stochastic amplitudes and phases~\cite{Gelmini:2020kcu, Yao:2024fie}. In the detector rest frame, the spatial components of the DP field are expressed as
\begin{equation}
    \label{eq:EMDP}
	\mathbf A'(\mathbf x,t)={\rm Re}\int\dd^3v\,\boldsymbol{\mathcal A}(\mathbf v)\exp\left[i\mathbf k_{\mathbf v}^{\rm cav}\cdot\mathbf x-i\omega_{\mathbf v}t\right]\,,
\end{equation}
where $\boldsymbol{\mathcal A}(\mathbf v)$ is the stochastic Fourier amplitude for a velocity component in cavity coordinates. For a fully polarized population, we write $\boldsymbol{\mathcal A}(\mathbf v) = \mathcal A(\mathbf v)\hat{\mathbf e}_A^{\rm cav}$, where $\hat{\mathbf e}_A^{\rm cav}$ is a polarization vector and $\mathcal A(\mathbf v)$ gives the stochastic amplitude and phase for each momentum component. Here, $\mathbf v$ is the velocity relative to the detector expressed in Galactic coordinates. In the nonrelativistic limit, the corresponding momentum and frequency in cavity coordinates are
\begin{align}
    \mathbf k_{\mathbf v}^{\rm cav} &= \mDP\mathcal R_{\rm gal\to cav}(t)\mathbf v\,,\\
    \omega_{\mathbf v} &\simeq \mDP\left(1+\frac{v^2}{2}\right)\,.
\end{align}

The orientation of the cavity and the local velocity distribution evolve on timescales much longer than the characteristic coherence time of the DP field. Their slow time evolution is kept when evaluating the signal modulation. Following the stochastic description of ultralight vector DM~\cite{Amaral:2024tjg,Yao:2024fie}, we assume that Fourier amplitudes associated with distinct velocities are statistically uncorrelated, with $\avg{\mathcal A_i(\mathbf v)}=0$ and $\avg{\mathcal A_i(\mathbf v)\mathcal A_j(\mathbf v')}=0$. The remaining two-point correlation function is parametrized as
\begin{equation}
	\label{eq:correlation}
	\avg{\mathcal A_i(\mathbf v)\mathcal A_j^*(\mathbf v')}=\frac{2\rho_\infty}{\mDP^2}F(\mathbf v,t)\Pi_{ij}^{\rm cav}(\mathbf v,t)\delta^{(3)}(\mathbf v-\mathbf v')\,,
\end{equation}
where $\rho_\infty$ is the asymptotic DPDM density outside the gravitational potential of the Solar System, $F(\mathbf v,t)$ is the local velocity distribution relative to this density, and $\Pi_{ij}^{\rm cav}(\mathbf v,t)$ is a Hermitian polarization density matrix in cavity coordinates. In general, the polarization density matrix can depend on both velocity and time. The polarization matrices in the Galactic and cavity frames are related through the rotation
\begin{equation}
    \Pi^{\rm cav}=\mathcal R_{\rm gal\to cav}\Pi^{\rm gal}\mathcal R_{\rm gal\to cav}^{T}\,.
\end{equation}
We consider a benchmark in which a fraction $\kappa_P$ of the DP energy density is in the polarized component, and the remaining fraction is distributed isotropically over the three spatial polarization components. The limits $\kappa_P=1$ and $\kappa_P=0$ correspond to a fully polarized and an isotropic unpolarized population, respectively. In Galactic coordinates, the corresponding polarization density matrix is
\begin{equation}
    \label{eq:polarization_matrix}
    \Pi_{ij}^{\rm gal}=\kappa_P e_{A,i}^{\rm gal}(e_{A,j}^{\rm gal})^*+\frac{1-\kappa_P}{3}\delta_{ij}\,.
\end{equation}
Since the isotropic component is invariant under spatial rotations, the polarization density matrix in the cavity frame is also described by an expression like in Eq.~\eqref{eq:polarization_matrix}, with the polarizations given by 
\begin{equation}
    \label{eq:polarization_vector_rotation}
    \hat{\mathbf e}_A^{\rm cav}(t)=\mathcal R_{\rm gal\to cav}(t)\hat{\mathbf e}_A^{\rm gal}\,.
\end{equation}
The rotation of the Earth induces a temporal variation of the polarized component in cavity coordinates, while leaving the isotropic component unchanged.

The normalization of Eq.~\eqref{eq:correlation} follows from the energy density of the DP field. At leading nonrelativistic order, using Eq.~\eqref{eq:EMDP}, we obtain
\begin{equation}
	\label{eq:DP_energy_density}
	\rho_{\rm loc}(t)\simeq\mDP^2\avg{\mathbf A'\cdot\mathbf A'}=\frac{\mDP^2}{2}\int\dd^3v\,\dd^3v’\,\avg{\boldsymbol{\mathcal A}(\mathbf v)\cdot\boldsymbol{\mathcal A}^{*}(\mathbf v')}\,.
\end{equation}
Substituting Eq.~\eqref{eq:correlation} and using the unit trace of the polarization density matrix gives
\begin{equation}
\label{eq:localdensity}
	\rho_{\rm loc}(t)=\rho_\infty\int\dd^3v\,F(\mathbf v,t)\,.
\end{equation}
In the absence of gravitational focusing, the velocity distribution is normalized to unity and $\rho_{\rm loc}=\rho_\infty$. In the presence of gravitational focusing, the integral of $F(\mathbf v,t)$ accounts for the gravitational enhancement of the local DP density.

As a benchmark for the Galactic DM population, we adopt the Standard Halo Model (SHM), in which the velocity distribution in the Galactic rest frame is described by a truncated Maxwellian distribution~\cite{Evans:2018bqy},
\begin{equation}
    \label{eq:shm}
    f_{\rm gal}(\mathbf u)=\frac{1}{(2\pi\sigma_v^2)^{3/2}N_{\rm esc}}\exp\left(-\frac{u^2}{2\sigma_v^2}\right)\Theta(v_{\rm esc}-u)\,,
\end{equation}
where $\mathbf u$ is the DM velocity relative to the Galactic rest frame, $\sigma_v$ is the one dimensional velocity dispersion, $v_{\rm esc}$ is the Galactic escape velocity, $\Theta$ is the Heaviside step function, and $N_{\rm esc}$ is the normalization factor. For our reference SHM, we adopt $\sigma_v=220/\sqrt{2}\,{\rm km\,s^{-1}}$ and $v_{\rm esc}=544\,{\rm km\,s^{-1}}$.

\section{Spectral response of a resonant haloscope}
\label{sec:haloscope}

The kinetic mixing between photons and DPs induces an effective EM current that can excite the resonant modes of a cavity even in the absence of an external magnetic field. To leading order in the kinetic mixing parameter $\chi$, the modified Maxwell equations in vacuum read
\begin{equation}
    \label{eq:modified_maxwell}
    \partial_\mu F^{\mu\nu}=\chi\mDP^2 A'^\nu\,,
\end{equation}
where $F^{\mu\nu}$ is the electromagnetic field strength tensor. The Proca constraint $\partial_\mu A'^\mu=0$ imposes the conservation of the four-current, with components 
\begin{equation}
    \rho_{\rm eff}=\chi\mDP^2 A'^0\,,\qquad \mathbf J_{\rm eff}=\chi\mDP^2\mathbf A'\,.
\end{equation}
For a nonrelativistic DP population, the time component of the field is suppressed with respect to its spatial components. Here, we consider the excitation of the resonant transverse EM modes of a vacuum cavity.

The EM response can be described through an expansion of the transverse component of the EM vector potential in the eigenmodes of the cavity~\cite{Rosenberg:2000wb,Hawton:2025kge,Jackson:1998nia},
\begin{equation}
    \mathbf A_{\rm T}(\mathbf x,t) = \sum_{\alpha} \boldsymbol\epsilon_{\alpha}(\mathbf x) \psi_{\alpha}(t)\,,
\end{equation}
where $\alpha \in \{m,n,p\}$ and the distinction between transverse magnetic (TM) and transverse electric (TE) modes. The spatial eigenfunctions $\boldsymbol\epsilon_{\alpha}(x)$ satisfy
\begin{equation}
\label{eq:cavity_eigenvalue}
\boldsymbol\nabla\times\left(\boldsymbol\nabla\times\boldsymbol\epsilon_\alpha\right)=\omega_\alpha^2\boldsymbol\epsilon_\alpha\,,\qquad \boldsymbol\nabla\cdot\boldsymbol\epsilon_\alpha=0\,,
\end{equation}
along with the boundary conditions of a perfectly conducting cavity with unit vector normal to the surface $\hat{\mathbf n}$ as $\hat{\mathbf n}\times\boldsymbol\epsilon_\alpha=0$ at the cavity walls. We adopt an orthonormal basis of real eigenfunctions over the volume of the cavity $V$ as
\begin{align}
\label{eq:norm}
    \int_V\dd^3x\,\boldsymbol\epsilon_\alpha(\mathbf x)\cdot\boldsymbol\epsilon_\beta(\mathbf x)=\delta_{\alpha\beta}\,.
\end{align}
For an ideal cylindrical cavity of radius $R$ and length $d$, the resonant frequencies of the TM and TE modes are
\begin{align}
    \omega_{{\rm TM}_{mnp}}^2&=\frac{x_{mn}^2}{R^2}+\frac{p^2\pi^2}{d^2}\,,\\
    \omega_{{\rm TE}_{mnp}}^2&=\frac{(x'_{mn})^2}{R^2}+\frac{p^2\pi^2}{d^2}\,,
\end{align}
where $x_{mn}$ and $x'_{mn}$ are the $n$-th positive roots of the Bessel function $J_m(x)$ and its derivative $J'_m(x)$, respectively. The damping coefficient accounting for the finite conductivity of the cavity walls and the coupling, in terms of the loaded quality factor of the mode $Q_\alpha$, is
\begin{equation}
    \label{eq:damping}
    \gamma_\alpha=\frac{\omega_\alpha}{Q_\alpha}\,.
\end{equation}
We assume that the dissipation can be described by independent damping coefficients for the cavity eigenmodes.

In the presence of the DP field, projecting the modified Maxwell equations onto the cavity eigenfunctions gives
\begin{equation}
\label{eq:driven_oscillator}
\ddot\psi_\alpha+\gamma_\alpha\dot\psi_\alpha+\omega_\alpha^2\psi_\alpha=\chi\mDP^2\int_V\dd^3x\,\boldsymbol\epsilon_\alpha(\mathbf x)\cdot\mathbf A'(\mathbf x,t)\,.
\end{equation}
The projection over the eigenfunctions removes the gradient contribution associated with the EM scalar potential under the conducting boundary conditions. We consider the stationary response after the transient solution of Eq.~\eqref{eq:driven_oscillator} has decayed, assuming that the cavity relaxation time is much shorter than the timescales associated with the variation of the detector orientation and the local velocity distribution. The solution for the EM mode amplitude is
\begin{equation}
    \label{eq:mode_solution}
    \psi_\alpha(t)=\chi\mDP^2\operatorname{Re}\int\dd^3v\,\frac{\exp(-i\omega_{\mathbf v}t)}{\omega_\alpha^2-\omega_{\mathbf v}^2-i\gamma_\alpha\omega_{\mathbf v}}\int_V\dd^3x\,\boldsymbol\epsilon_\alpha(\mathbf x)\cdot\boldsymbol{\mathcal A}(\mathbf v)\exp(i\mathbf k_{\mathbf v}^{\rm cav}\cdot\mathbf x)\,,
\end{equation}
where the momentum $\mathbf k_{\mathbf v}^{\rm cav}$ and the frequency $\omega_{\mathbf v}$ are defined in Sec.~\ref{sec:formfactor}.

The coupling between the DP field and a given EM cavity mode is defined in terms of the dimensionless vector overlap
\begin{equation}
	\label{eq:vector_overlap}
	\mathbf I_\alpha(\mathbf k)=\frac{\displaystyle\int_V\dd^3x\,\mathbf E_\alpha^*(\mathbf x)\exp(i\mathbf k\cdot\mathbf x)}{\displaystyle\sqrt{V\int_V\dd^3x\,|\mathbf E_\alpha(\mathbf x)|^2}}=\frac{1}{\sqrt V}\int_V\dd^3x\,\boldsymbol\epsilon_\alpha(\mathbf x)\exp(i\mathbf k\cdot\mathbf x)\,,
\end{equation}
where $\mathbf E_\alpha(\mathbf x)$ is the spatial electric field profile of the cavity mode and in the last step we adopted the orthonormal eigenfunctions in Eq.~\eqref{eq:norm}. For a fully polarized DP population with polarization vector $\hat{\mathbf e}_A^{\rm cav}$, the dimensionless form factor describing the spatial overlap between the electric field of the cavity mode and the effective current induced by the DP field is
\begin{equation}
    \label{eq:formfactor_polarized}
    C_\alpha(\mathbf k,\hat{\mathbf e}_A^{\rm cav})=\left|\mathbf I_\alpha(\mathbf k)\cdot\hat{\mathbf e}_A^{\rm cav}\right|^2=\frac{\displaystyle\left|\int_V\dd^3x\,\mathbf E_\alpha^*(\mathbf x)\cdot\hat{\mathbf e}_A^{\rm cav}\exp(i\mathbf k\cdot\mathbf x)\right|^2}{\displaystyle V\int_V\dd^3x\,|\mathbf E_\alpha(\mathbf x)|^2}\,.
\end{equation}
Note, that we account for both the polarization and the spatial variation of the DP field across the cavity. For a general polarization state described by the density matrix $\Pi^{\rm cav}$ introduced in Sec.~\ref{sec:formfactor}, the form factor generalizes to
\begin{equation}
    \label{eq:formfactor_general}
    C_\alpha(\mathbf k,\Pi^{\rm cav})=\sum_{i,j}I_{\alpha i}(\mathbf k)\Pi_{ij}^{\rm cav}I_{\alpha j}^*(\mathbf k)\,.
\end{equation}
For the polarization matrix in Eq.~\eqref{eq:polarization_matrix}, expressed in terms of the labortatory frame, gives 
\begin{equation}
    \label{eq:formfactor_partial}
    C_\alpha(\mathbf k,t)=\kappa_P\left|\mathbf I_\alpha(\mathbf k)\cdot\hat{\mathbf e}_A^{\rm cav}(t)\right|^2+\frac{1-\kappa_P}{3}|\mathbf I_\alpha(\mathbf k)|^2\,,
\end{equation}
where the first term describes the contribution from the population with a common polarization state in Galactic coordinates, and the second term is the contribution from the isotropically polarized population. In the limit in which the spatial variation of the DP field across the cavity of size $\ell_{\rm cav}$ can be neglected, $|\mathbf k|\ell_{\rm cav}\ll1$, the vector overlap reduces to
\begin{equation}
	\label{eq:overlap_homogeneous}
	\mathbf I_\alpha(0)=\frac{\displaystyle\int_V\dd^3x\,\mathbf E_\alpha^*(\mathbf x)}{\displaystyle\sqrt{V\int_V\dd^3x\,|\mathbf E_\alpha(\mathbf x)|^2}}\,.
\end{equation}
In this limit, the geometrical form factor is independent of the DP momentum distribution, although its dependence on the polarization state and the orientation of the cavity is preserved. The time dependence comes from the rotation of the polarization vector, while the contribution from the isotropic component is independent of it.

For a general momentum distribution, we define the velocity averaged form factor
\begin{equation}
\label{eq:averaged_formfactor}
\overline C_\alpha(t)=\frac{\displaystyle\int\dd^3v\,F(\mathbf v,t)C_\alpha\left(\mathbf k_{\mathbf v}^{\rm cav},\Pi^{\rm cav}(\mathbf v,t)\right)}{\displaystyle\int\dd^3v\,F(\mathbf v,t)}\,,
\end{equation}
where the normalization in the denominator reminds us that the local velocity distribution $F(\mathbf v,t)$ might not integrate to one in the presence of gravitational focusing~\cite{Alenazi:2006wu, Kim:2021yyo}.

The energy associated with the EM oscillator is
\begin{equation}
	\label{eq:mode_energy}
	U_\alpha(t)=\frac{1}{2}\left[\dot\psi_\alpha^2(t)+\omega_\alpha^2\psi_\alpha^2(t)\right]\,,
\end{equation}
so that the corresponding energy loss away from resonance reads $P_\alpha(t)=\gamma_\alpha\dot\psi_\alpha^2(t)$. For a monochromatic field at resonance, we recover $\avg{P_\alpha}=\gamma_\alpha\avg{U_\alpha}$. Using Eq.~\eqref{eq:mode_solution} and the definition of the vector overlap in Eq.~\eqref{eq:vector_overlap_normalized}, the ensemble averaged power deposited in the cavity mode is
\begin{equation}
\label{eq:power}
\avg{P_\alpha(t)}=\chi^2\rho_\infty V\mDP^2\int\dd^3v\,F(\mathbf v,t)\frac{\gamma_\alpha\omega_{\mathbf v}^2}{(\omega_\alpha^2-\omega_{\mathbf v}^2)^2+\gamma_\alpha^2\omega_{\mathbf v}^2}C_\alpha\left(\mathbf k_{\mathbf v}^{\rm cav},\Pi^{\rm cav}(\mathbf v,t)\right)\,.
\end{equation}
Equation~\eqref{eq:power} represents one of the main results of this work, and describes the response of an EM cavity mode to a stochastic DP field with an arbitrary nonrelativistic momentum distribution and polarization density matrix. This expression includes the spatial variation of the DP field across the cavity, the frequency dependence of the EM response, and the time dependence associated with the motion of the detector. The conventional expression for the resonant haloscope power is recovered when the cavity is tuned to the DP signal and its linewidth exceeds the characteristic width of the DP spectrum, in the regime $Q_\alpha\ll\sigma_v^{-2}$ where
\begin{equation}
	\frac{\gamma_\alpha\omega_{\mathbf v}^2}{(\omega_\alpha^2-\omega_{\mathbf v}^2)^2+\gamma_\alpha^2\omega_{\mathbf v}^2}\simeq\frac{1}{\gamma_\alpha}\simeq\frac{Q_\alpha}{\mDP}\,.
\end{equation}
Using the definition of the local DP density in Eq.~\eqref{eq:localdensity} and the velocity averaged form factor in Eq.~\eqref{eq:averaged_formfactor}, Eq.~\eqref{eq:power} becomes
\begin{equation}
\label{eq:resonant_power}
\avg{P_\alpha(t)}\simeq\chi^2\rho_{\rm loc}(t)V\mDP Q_\alpha\overline C_\alpha(t)\,.
\end{equation}

\section{Results}
\label{sec:results}

We apply the formalism developed above to two cylindrical cavity configurations. The \emph{prototype} FLASH cavity has its cylinder axis perpendicular to the Earth's surface, while the second configuration corresponds to the FLASH cavity, whose axis lies in the horizontal plane and forms an angle $\theta$ with the local East direction towards the North. The dimensions of the two cavities are reported in Table~\ref{tab:cavity_dimensions}. For FLASH, we adopt the reference orientation $\theta=66^\circ$. For cavities at the same geographical location, with the prototype axes pointing towards the local South, East and zenith, respectively, the transformation between the prototype and FLASH cavity frames is described by the rotation matrix
\begin{equation}
\label{eq:prototype_flash_rotation}
\mathcal R_{\rm proto\to FLASH}=
\begin{pmatrix}
\cos\theta & \sin\theta & 0\\
0 & 0 & -1\\
-\sin\theta & \cos\theta & 0
\end{pmatrix}\,.
\end{equation}
Both the polarization vector and the momentum of the DP field transform according to the same rotation at leading nonrelativistic order. The choice of transverse cavity axes fixes the orientation of the two degenerate TE modes.

\begin{table}[htbp]
\centering
\begin{tabular}{lcc}
\hline\hline
 & Prototype & FLASH \\
\hline
Radius $R$ (mm) & 175.25 & 950 \\
Length $d$ (mm) & 276 & 1500 \\
Volume $V$ (${\rm m}^3$) & 0.0266 & 4.253 \\
Aspect ratio $R/d$ & 0.635 & 0.633 \\
\hline\hline
\end{tabular}
\caption{Geometrical parameters of the prototype and FLASH cavities considered in this work.}
\label{tab:cavity_dimensions}
\end{table}

In the homogeneous field limit, the cylindrical cavity modes with nonvanishing overlaps belong to the $\mathrm{TM}_{0n0}$ and $\mathrm{TE}_{1np}$ families, with odd $p$ for the latter. We denote the maximum homogeneous field form factor of a mode by $C_\alpha^{\rm max}$. For the TM modes and the two degenerate TE modes, whose integrated electric fields are taken along the transverse cavity axes, Eq.~\eqref{eq:formfactor_general} gives
\begin{align}
C_{\mathrm{TM}_{0n0}}(0,t)&=\frac{4}{x_{0n}^2}\Pi_{zz}^{\rm cav}(t)\,,\\
C_{\mathrm{TE}_{1np}^{(a)}}(0,t)&=\frac{16}{p^2\pi^2[(x'_{1n})^2-1]}\Pi_{xx}^{\rm cav}(t)\,,\\
C_{\mathrm{TE}_{1np}^{(b)}}(0,t)&=\frac{16}{p^2\pi^2[(x'_{1n})^2-1]}\Pi_{yy}^{\rm cav}(t)\,.
\label{eq:cylindrical_formfactors}
\end{align}
The prefactors follow by integrating the normalized electric field profiles over the cavity volume. The TM modes probe the axial polarization component, whereas the two TE modes probe orthogonal transverse components. For the fundamental modes, $C_{\mathrm{TM}_{010}}^{\rm max}=0.69166$ and $C_{\mathrm{TE}_{111}}^{\rm max}=0.67831$. The form factors for higher modes decrease with increasing radial or axial mode number. Since the trace of the polarization matrix is unity, the sum of the three orthogonal responses, each divided by its maximum form factor, is independent of the polarization state. The corresponding modes generally resonate at different frequencies.

For the $\mathrm{TM}_{0n0}$ modes, the finite momentum overlap can also be evaluated analytically using the electric field profile proportional to $J_0(x_{0n}r_\perp/R)\hat{\mathbf z}$. Defining $q=k_\perp R$ and $\operatorname{sinc}(x)=\sin(x)/x$, we obtain
\begin{equation}
\label{eq:TM_finite_momentum}
C_{\mathrm{TM}_{0n0}}(\mathbf k,t)=\frac{4x_{0n}^2J_0^2(q)}{(x_{0n}^2-q^2)^2}\operatorname{sinc}^2\left(\frac{k_zd}{2}\right)\Pi_{zz}^{\rm cav}(\mathbf v,t)\,.
\end{equation}
The homogeneous field result is recovered for $q,k_zd\to0$. For the fundamental modes at resonance and typical Galactic velocities, the finite spatial phase corrections are small, of order $v^2$. The results below focus on the temporal dependence associated with polarization rather than on the small corrections associated with the spatial phase.

We also consider the wall losses associated with the different modes. The intrinsic quality factor of a cavity with uniform surface resistance $R_s(\omega_\alpha)$ is $Q_{\alpha,0}=G_\alpha/R_s(\omega_\alpha)$, where the geometrical factor in SI units is
\begin{equation}
G_\alpha=\mu_0\omega_\alpha\frac{\int_V\dd^3x\,|\mathbf H_\alpha|^2}{\int_{\partial V}\dd S\,|\mathbf H_{\alpha,t}|^2}\,.
\end{equation}
For the modes of interest, defining $a=x'_{1n}$, $b=p\pi R/d$ and the vacuum impedance $\eta_0$, the geometrical factors are
\begin{align}
G_{\mathrm{TM}_{0n0}}&=\frac{\eta_0 x_{0n}}{2(1+R/d)}\,,\\
G_{\mathrm{TE}_{1np}}&=\frac{\eta_0}{2}\frac{\sqrt{a^2+b^2}(1-a^{-2})(1+b^2/a^2)}{1+b^2/a^4+2(R/d)(1-a^{-2})b^2/a^2}\,.
\end{align}
The two cavities have nearly identical aspect ratios and hence similar geometrical factors for corresponding modes. Their resonant frequencies differ because of their different dimensions. The frequencies and maximum form factors of the modes considered in the following figures, together with the geometrical factors for FLASH, are reported in Table~\ref{tab:cavity_modes}. The loaded quality factor entering Eq.~\eqref{eq:power} also depends on the external coupling.

\begin{table}[htbp]
\centering
\begin{tabular}{lrrrr}
\hline\hline
Mode & $f_{\rm proto}$ (MHz) & $f_{\rm FLASH}$ (MHz) & $C_\alpha^{\rm max}$ & $G_{\rm FLASH}$ ($\Omega$) \\
\hline
$\mathrm{TM}_{010}$ & 654.74 & 120.78 & 0.69166 & 277.34 \\
$\mathrm{TM}_{020}$ & 1502.89 & 277.24 & 0.13127 & 636.61 \\
$\mathrm{TM}_{030}$ & 2356.06 & 434.63 & 0.05341 & 998.00 \\
$\mathrm{TM}_{040}$ & 3210.36 & 592.23 & 0.02877 & 1359.87 \\
$\mathrm{TM}_{050}$ & 4065.08 & 749.90 & 0.01794 & 1721.92 \\
$\mathrm{TM}_{060}$ & 4920.02 & 907.61 & 0.01225 & 2084.06 \\
$\mathrm{TE}_{111}$ & 739.08 & 136.15 & 0.67831 & 326.89 \\
$\mathrm{TE}_{113}$ & 1704.68 & 313.73 & 0.07537 & 708.00 \\
$\mathrm{TE}_{115}$ & 2761.39 & 508.14 & 0.02713 & 1136.91 \\
$\mathrm{TE}_{121}$ & 1549.81 & 285.81 & 0.05911 & 1002.67 \\
$\mathrm{TE}_{131}$ & 2386.71 & 440.23 & 0.02256 & 1606.61 \\
$\mathrm{TE}_{141}$ & 3233.01 & 596.36 & 0.01192 & 2203.92 \\
$\mathrm{TE}_{151}$ & 4083.03 & 753.18 & 0.00737 & 2798.89 \\
$\mathrm{TE}_{161}$ & 4934.87 & 910.33 & 0.00501 & 3392.74 \\
\hline\hline
\end{tabular}
\caption{Resonant frequencies, maximum homogeneous field form factors and wall loss geometrical factors for the ideal cylindrical cavity modes. The maximum form factors refer to optimal polarization alignment with an individual mode. Both orientations of each degenerate TE mode have the same maximum form factor and geometrical factor.}
\label{tab:cavity_modes}
\end{table}

\begin{figure}[H]
    \centering
    \begin{subfigure}{0.48\textwidth}
        \centering
        \includegraphics[width=\textwidth]{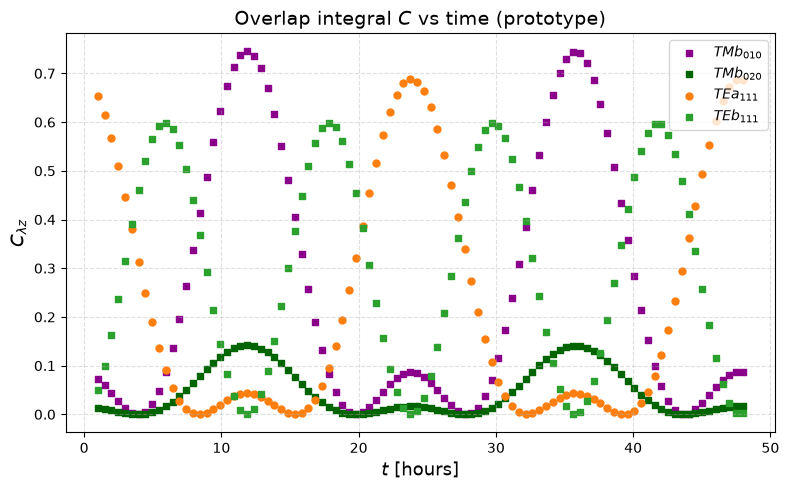}
    \end{subfigure}
    \hfill
    \begin{subfigure}{0.48\textwidth}
        \includegraphics[width=\textwidth]{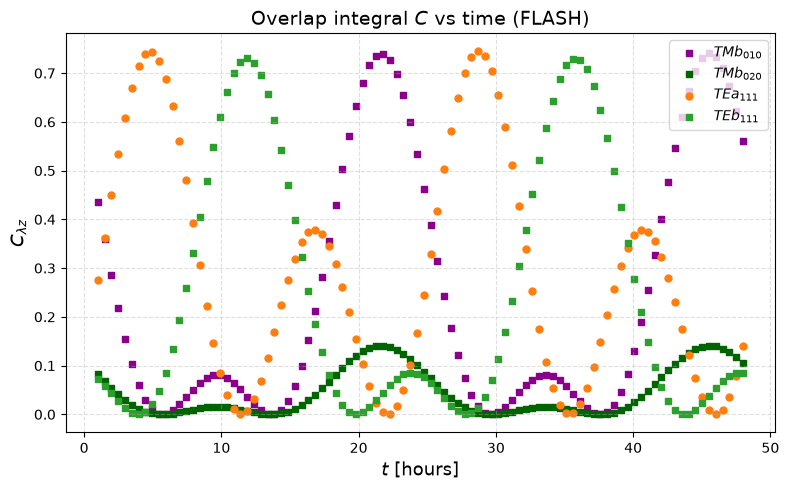}
        \centering
    \end{subfigure}
    \caption{Numerical overlap estimator $\widetilde C_\alpha$ over approximately two sidereal days for a DP field linearly polarized along the Galactic $z$ axis. The prototype (left) and FLASH (right) configurations are shown for the $\mathrm{TM}_{010}$, $\mathrm{TM}_{020}$ and two degenerate $\mathrm{TE}_{111}$ modes. The plotted estimator illustrates the temporal profiles; its absolute normalization is not used in the physical power calculation.}
\label{fig:set1modes}
\end{figure}

\begin{figure}[H]
    \centering
    \begin{subfigure}{0.48\textwidth}
        \centering
        \includegraphics[width=\textwidth]{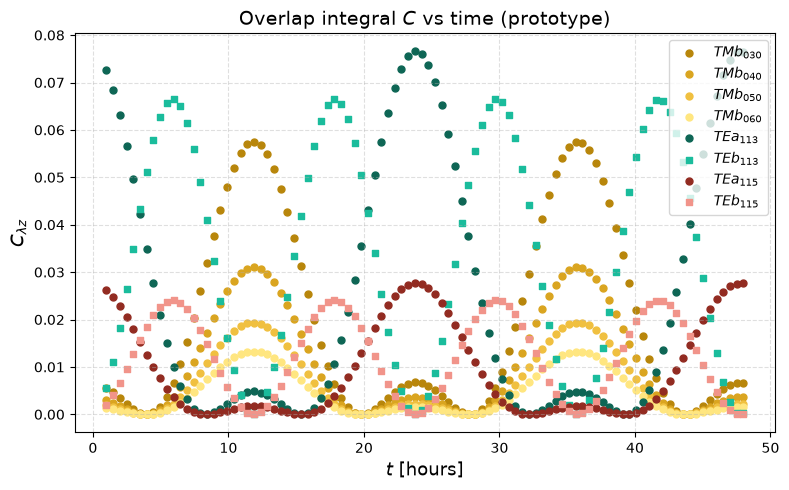}
    \end{subfigure}
    \hfill
    \begin{subfigure}{0.48\textwidth}
        \centering
        \includegraphics[width=\textwidth]{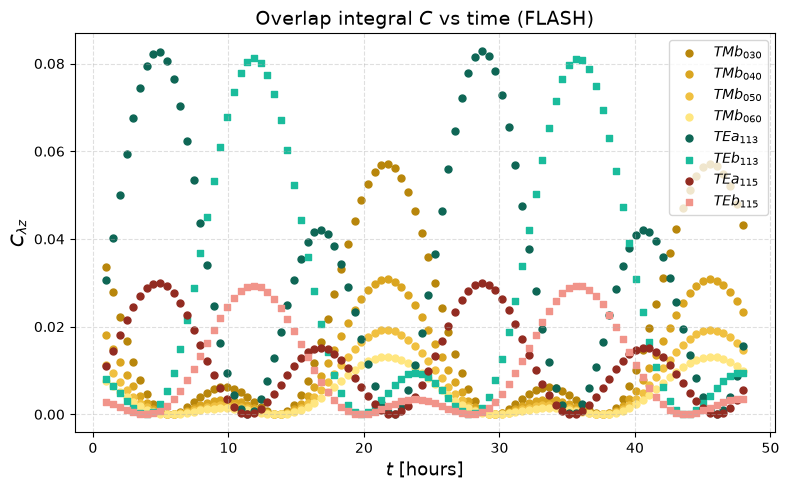}
    \end{subfigure}
    \caption{As in Fig.~\ref{fig:set1modes}, for the higher radial $\mathrm{TM}_{0n0}$ modes with $n=3,4,5,6$ and the two degenerate orientations of the $\mathrm{TE}_{113}$ and $\mathrm{TE}_{115}$ modes.}
\label{fig:set2modes}
\end{figure}

\begin{figure}[H]
    \centering
    \begin{subfigure}{0.48\textwidth}
        \centering
        \includegraphics[width=\textwidth]{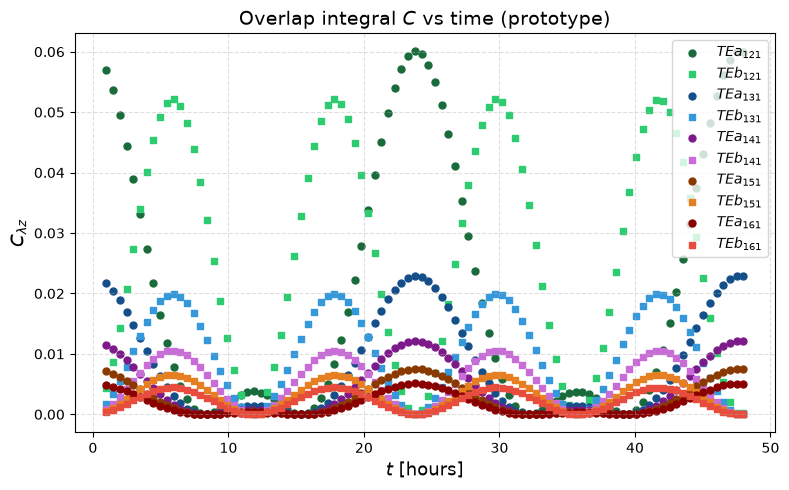}
    \end{subfigure}
    \hfill
    \begin{subfigure}{0.48\textwidth}
        \centering
        \includegraphics[width=\textwidth]{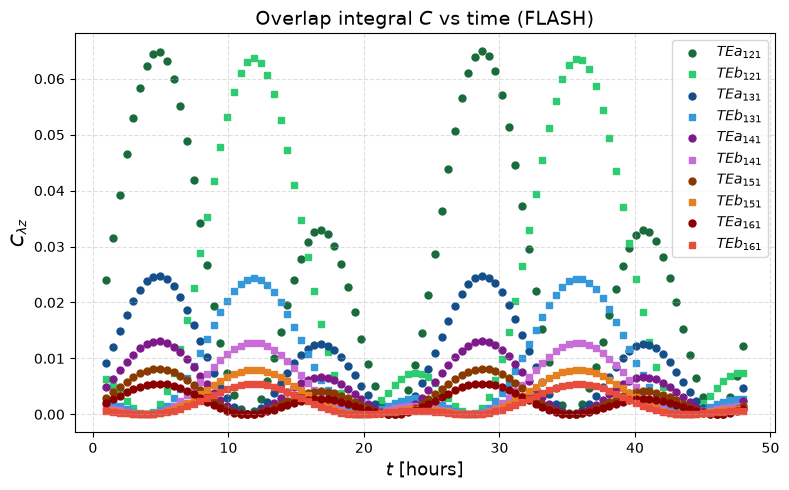}
    \end{subfigure}
    \caption{As in Fig.~\ref{fig:set1modes}, for both orientations of the $\mathrm{TE}_{1n1}$ modes with $n=2,3,4,5,6$.}
\label{fig:set3modes}
\end{figure}

Figures~\ref{fig:set1modes}--\ref{fig:set3modes} illustrate the temporal modulation for a fully polarized DP population aligned with the Galactic $z$ axis. We retain the numerical overlap estimator $\widetilde C_\alpha(t)$ from the initial calculation to illustrate the modulation profiles. Its absolute normalization differs from that of the physical form factor in Eq.~\eqref{eq:formfactor_general}; the curves are not used to determine absolute signal powers or calibrated daily averages. The physical normalization is instead specified by the analytical overlaps and Table~\ref{tab:cavity_modes}. The agreement of the numerical estimator with the analytical directional profiles must be checked before interpreting it quantitatively.

Figure~\ref{fig:set1modes} illustrates the response of the fundamental axial and transverse modes, together with the $\mathrm{TM}_{020}$ mode. The axial and transverse responses exhibit maxima at different times, reflecting their sensitivity to different components of the DP polarization. The two degenerate TE modes have complementary responses, so that reading out both modes can reduce the suppression associated with an unfavorable alignment with either individual mode. Their combined response remains insensitive to a fully polarized field parallel to the cylinder axis in the homogeneous field limit.

Figures~\ref{fig:set2modes} and \ref{fig:set3modes} extend this comparison to higher radial and axial modes. The smaller overlaps reflect the cancellations between regions of the cavity in which the electric field has opposite signs. Within each family and for a fixed mode orientation, the homogeneous field response has the same temporal dependence, multiplied by a mode dependent geometrical factor. This accounts for the similar temporal profiles and progressively smaller overlaps of the higher modes in these figures. The geometrical suppression does not by itself determine the relative signal powers, which also depend on the resonance frequencies and loaded quality factors.

\begin{figure}[H]
    \centering
    \begin{subfigure}{0.48\textwidth}
        \centering
        \includegraphics[width=\textwidth]{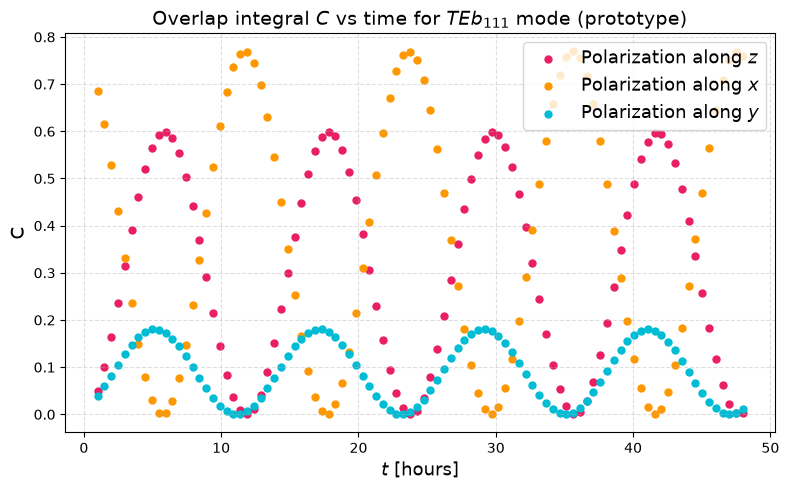}
    \end{subfigure}
    \hfill
    \begin{subfigure}{0.48\textwidth}
        \centering
        \includegraphics[width=\textwidth]{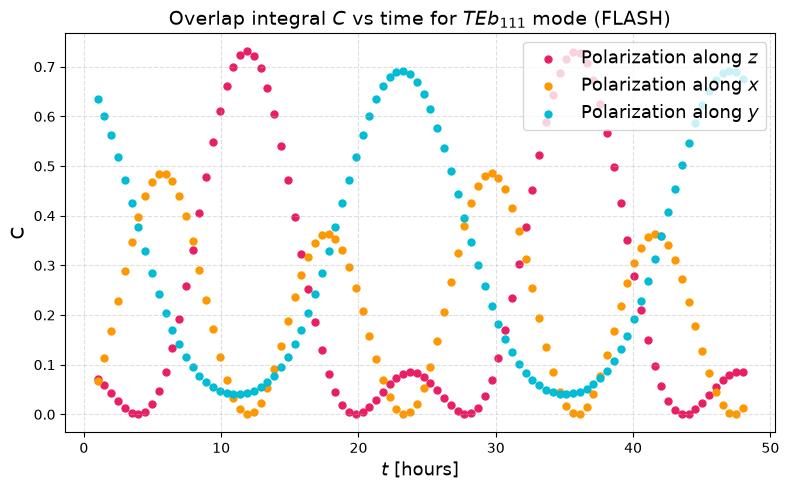}
    \end{subfigure}
    \caption{Numerical overlap estimator for the $\mathrm{TE}_{111}^{(b)}$ mode for a linearly polarized DP field aligned with the Galactic $x$, $y$ or $z$ axis. The prototype (left) and FLASH (right) configurations are shown. The curves illustrate the dependence of the temporal profile on the assumed polarization.}
\label{fig:comparemodes}
\end{figure}

The temporal profiles also depend on the Galactic polarization direction. Figure~\ref{fig:comparemodes} illustrates this dependence for the $\mathrm{TE}_{111}^{(b)}$ mode and the three benchmark polarizations along the Galactic Cartesian axes. Different polarization directions lead to different modulation amplitudes and phases. These differences can in principle provide information on the polarization of a detected signal, although a single mode does not generally determine an arbitrary polarization state uniquely.

\begin{figure}[H]
    \centering
    \includegraphics[width=0.8\linewidth]{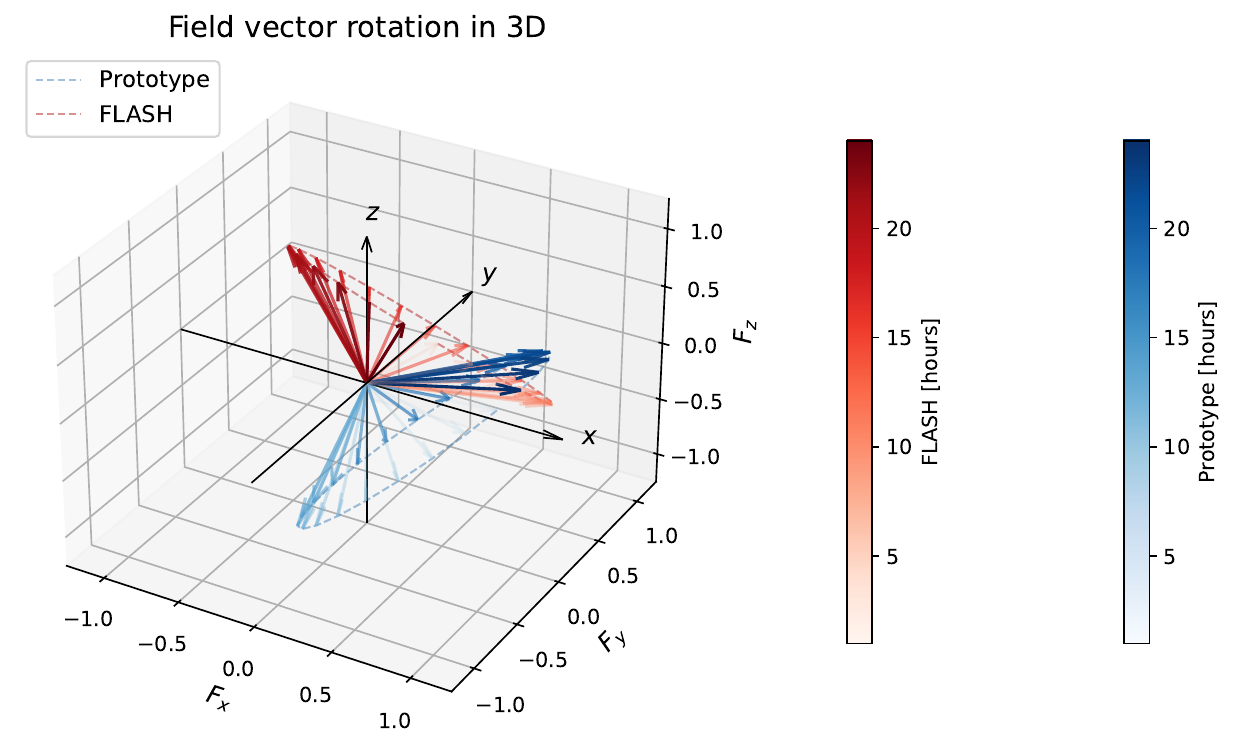}
    \caption{Illustration of the evolution of the components $F_x,F_y,F_z$ of a field polarized along the Galactic $z$ axis, expressed in the detector frames of the prototype (blue) and FLASH (red) configurations. The color scale indicates the elapsed time in hours, rather than a change in the field magnitude.}
\label{fig:rotation}
\end{figure}

The approximately shifted temporal profiles of the TM modes in Fig.~\ref{fig:set1modes} can be understood from the projection of the cavity axis onto the Earth's rotation axis. For a linearly polarized field, defining $\mu=\hat{\mathbf e}_A^{\rm gal}\cdot\hat{\mathbf s}$ and $h_\alpha=\hat{\mathbf n}_\alpha\cdot\hat{\mathbf s}$, where $\hat{\mathbf s}$ is the Earth's rotation axis and $\hat{\mathbf n}_\alpha$ is the direction of the integrated electric field, the polarization projection is
\begin{equation}
\hat{\mathbf e}_A^{\rm gal}\cdot\hat{\mathbf n}_\alpha(t)=\mu h_\alpha+\sqrt{(1-\mu^2)(1-h_\alpha^2)}\cos(\Omega_\oplus t+\varphi_\alpha)\,.
\end{equation}
Here $\Omega_\oplus$ is the sidereal angular frequency and $\varphi_\alpha$ is an orientation dependent phase. At geographical latitude $\lambda$, the prototype and FLASH cylinder axes have projections $h_{\rm proto}=\sin\lambda$ and $h_{\rm FLASH}=\sin\theta\cos\lambda$, respectively. For the reference latitude $\lambda=41.82^\circ$ and $\theta=66^\circ$, these projections are approximately $0.667$ and $0.681$. This explains why the corresponding TM modulation profiles can have similar shapes with a relative phase shift. In general, changing the cavity orientation modifies both the modulation amplitude and the phase. The evolution of the field components in the two detector frames is illustrated in Fig.~\ref{fig:rotation}.

\section{Discussion and conclusions}
\label{sec:discussion}

In this work, we have developed a formulation of the EM response of a resonant cavity to a stochastic DP field characterized by a general momentum distribution and polarization density matrix. The resulting spectral power accounts for the finite linewidth of the DP signal, the frequency dependence of the cavity response, and the spatial variation of the DP field across the detector. The motion of the detector through the Galactic halo and solar gravitational focusing enter through the local velocity distribution, while the rotation of the Earth modulates the response to a preferred Galactic polarization direction.

An important result of our analysis is the complementary response of the TM and TE modes of a cylindrical cavity. The maximum homogeneous field form factors of the fundamental $\TM_{010}$ and $\TE_{111}$ modes differ by less than two percent, demonstrating that transverse modes can provide a geometrical coupling to DPDM comparable to that of conventional axion haloscope modes. The two degenerate orientations of the $\TE_{111}$ mode probe orthogonal polarization components perpendicular to the cylinder axis. Combining the normalized responses of three orthogonal modes yields a polarization independent geometrical response in the homogeneous field limit. Exploiting this complementarity at a common DP mass, however, requires suitable tuning mechanisms, modifications of the cavity geometry, or independent resonators. The sidereal modulation of the signal gives an additional observable for investigating the polarization properties of the DP population. A preferred Galactic polarization direction generates contributions at the fundamental and second harmonics of the sidereal frequency, with relative amplitudes determined by the polarization and detector orientation. Measurements involving complementary EM modes can help constrain the polarization composition of a potential DP signal and discriminate it from instrumental backgrounds.

The spectral response provides complementary information on the local DP population. In the homogeneous field limit and for a polarization density matrix independent of velocity, the polarization dependence factorizes from the spectral dependence associated with the local velocity distribution. A preferred polarization direction then modulates the signal amplitude without modifying its instantaneous normalized spectral shape. This factorization is generally lost in the presence of correlations between momentum and polarization or when finite momentum effects become relevant. While spatial corrections are typically negligible for the fundamental modes of conventional haloscopes, they can become relevant for higher order modes, modes with vanishing homogeneous field overlap, and detectors extending over a significant fraction of the DP de Broglie wavelength. A systematic description of these corrections at second order in the DM velocity also requires accounting for the corresponding relativistic corrections to the polarization covariance and energy density.

Our results motivate the investigation of transverse EM modes in existing and planned haloscopes, including FLASH. The combination of complementary polarization responses, spectral measurements, and sidereal modulation can extend the scientific potential of these experiments beyond DM detection, providing access to the polarization and kinematic properties of the Galactic vector DM population.

\section*{Acknowledgments}
We thank Paolo Beltrame, Claudio Gatti, and Lei Wu for stimulating discussions. This work is supported by the Istituto Nazionale di Fisica Nucleare (INFN) through the Commissione Scientifica Nazionale~2 (CSN2), Esperimento QUAX, Virgo, Einstein Telescope, and the Commissione Scientifica Nazionale~4 (CSN4), Iniziativa Specifica ``Quantum Universe'' (QGSKY).

\appendix

\section{Dark Photon field of galactic origin in the detector reference frame}

We first define the reference frame which will enter the analysis. 
\begin{itemize}
    \item Galactic reference frame \\ $${x_{gal},y_{gal},z_{gal}}:$$ the $x$ points to the Galactic center, $y$ points in the direction of the Galactic rotation and $z$ is perpendicular to the plane defined by $x$ and $y$. 
    \item elliptic reference frame $${x_{hel}, y_{hel}, z_{hel}}:$$ $$x_{hel}$$ points in the direction of the vernal equinox and the $$y_{hel}$$ points in the direction of the summer solstice \\ 
    \item equatorial reference frame  $${x_{eq}, y_{eq}, z_{eq}}$$ which is titled wrt to the elliptic reference frame, given that it is built on the celestial equator, that represents the projection of the terrestrial equator on the celestial sphere and on the celestial pole which lies on an axis parallel to the axis rotation \\
    \item detector reference frame: $${x_{det}, y_{det}, z_{det}},$$ where $z_{det}$ axis is parallel to the axis of the cylinder and the $y_{det}$ axis points to the local east while $x_{det}$ points to the south (when a prototype-like cavity is considered).\\ 
\end{itemize}
We expect the shape of the DP field to be correlated to the dynamic of the Galaxy, if the field is characterized by a fixed polarization, three relevant cases will be the alignments with the $x_{gal},y_{gal},z_{gal}$. In the detector reference frame the field will have a different form, dictated by the change of reference frame made on the filed. In particular, if the new coordinates are rotated with respect to the old ones: $\vec{x}'= R \ \vec{x} $, then a vector field $\vec{\mathcal{F}}(x)$ will transform as:
\begin{equation}
    \vec{\mathcal{F}'}(x') =  R \ \mathcal{F}(R^{-1}\vec{x}') 
\end{equation}
In our case the the change of coordinates will be the results of two rotations: the first one links the galactic coordinates to the heliocentric ones, while the second one links the heliocentri coordiates to the detector reference frame. \\
First of all we define rectangula ecliptic geocentric coordinates from \cite{McCabe:2013kea}:
\begin{equation}
    x_{geo-ecl} = (R^{-1}\cdot P \cdot M^{-1}) \ x_{gal} \equiv \mathcal{R}_{1} x_{gal},
\end{equation}
where $M^{-1}$ transforms galactic coordinates into the equatorial coordinates at J2000.0, $P$ transforms the equatorial coordinates at j2000.0 to the equatorial coordinates at epoch of date T and finally $R^{-1}$ transforms the equatorial coordinates into the ecliptic-geocentric one. From a physical point of view, $M^{-1}$ takes into account the effect of the orientation of the Earth with respect to the centre of the Galaxy, $P$ models the effects of the precession of equinoxes and the torque on Earth caused by the external gravitational field and finally $R$ takes into account the effect of the inclination of the spin axis of the Earth. The explicit expressions for the matrices can be found in Appendix A and Appendix B of \cite{McCabe:2013kea}.\\
The second transformation bring the system from the heliocentric reference frame to the detector one:
\begin{equation}
    x_{detector} = \mathcal{R}_{2} \ x_{geo-ecl},
\end{equation}
with:
\begin{align}
    \mathcal{R}_{2} = \begin{bmatrix}
        \cos{\theta}\cos{\omega t} &  \cos{\theta}\sin{\omega t} & -\sin{\theta} \\
        -\sin{\omega t} & \cos{\omega t} & 0 \\
        \sin{\theta}\cos{\omega t} & \sin{\theta}\sin{\omega t} & \cos{\theta} \\
    \end{bmatrix},
\end{align}
where $\theta$ is the colatitude of the position of the cavity, while $\omega$ is the angular frequency of rotation of the Earth. \\
We obtain that:
\begin{equation}
    \vec{\mathcal{F}'}(x') =  \mathcal{R}_{2} \mathcal{R}_{1} \ \mathcal{F}(\mathcal{R}^{-1}_{1} \ \mathcal{R}^{-1}_{2}\vec{x}') \xrightarrow{unifrom \ field}  \mathcal{R}_{2} \mathcal{R}_{1} \ \mathcal{F}(\vec{x}')
\end{equation}

\section{Numerical evaluation of the overlap function}
\label{sec:numerical}

The computation of $$\mathcal C =  \int_{V} \ d^{3}x \ \epsilon(x) \ \Vec{\epsilon_{\alpha}}(x) \cdot (\sqrt{m_{A'}} \ \Vec{A'}(x))$$
proceeds in two steps: the computation of $A'(x)$ through an integration in the space of momenta and the integration in the physical space inside the cavity which is explicit in the formula above. \\
The field $A'(x)$ reads:
\begin{align}\label{eq:Afield}
   &A'(x) = \frac{ \ \sqrt{2 \rho_{DM}V}}{(2 \pi)^{3} m_{A'} } \ \sum_{\lambda} \int d^{3} k  \sqrt{\frac{P(k)}{\epsilon(x)}}\ \epsilon^{\mu}(\Vec{k}, \lambda)  e^{-ik \cdot x} \equiv \frac{\ \sqrt{2 \rho_{DM}V}}{(2 \pi )^{3} m_{A'}}  \ \sum_{\lambda} \int d^{3} k  \ \ A'(|k|, \theta_{k}, \phi_{k}) \ e^{ik \cdot x}\\
   &\epsilon^{\mu}(k, \lambda) = L^{\mu}_{\nu}(k) \ \epsilon^{\nu}(0,\lambda), \  \ \ \ \ \epsilon(0,0) = \begin{pmatrix}
           0 \\
           0 \\
           1 \\
           0
         \end{pmatrix}, \ \ \ \
\epsilon(0,1) = \frac{1}{\sqrt{2}}\begin{pmatrix}
           1 \\
           i \\
           0 \\
           0
         \end{pmatrix}, \ \ \ \ 
\epsilon(0,-1) = \frac{1}{\sqrt{2}}\begin{pmatrix}
           1 \\
           -i \\
           0 \\
           0
         \end{pmatrix}.
\end{align}
To integrate Eq. \ref{eq:Afield}, the angular and radial components of $\intk$ are separated. \\
The phase $e^{ik \cdot x}$ is expanded through the Rayleigh formula \cite{Mehrem:2009ip}: 
\begin{align}
    e^{i  k \cdot x} = 4 \pi \sum_{l'=0}^{\infty} \sum_{m'=-l'}^{l'} \ i^{l'} \  J_{l'}(|k| r) \ Y_{l',m'}^{*}(\theta_{k}, \phi_{k}) \ Y_{l',m'}(\theta_{x}, \phi_{x}),
\end{align}
where $Y$ are the spherical harmonic functions and $\bessel$ are the spherical Bessel functions which can be related to the Bessel functions of the first kind $J_{v}$ through:
\begin{equation}
    \bessel = \sqrt{\frac{\pi}{2 |k| r}} \ J_{v}(l'+\frac{1}{2}, |k|r). 
\end{equation}
Defining the angular coefficients $A_{lm}$
\begin{align}\label{eq:angularintegration}
 A_{lm} = \int_{\Omega_{k}} d \Omega_{k} \intk Y^{*}_{lm}(\theta_{k}, \phi_{k}).
\end{align}
and making use of the orthogonality and completeness relations of the spherical harmonic functions:
\begin{align}
    &\int_{\Omega_{k}} \ d \Omega_{k} \ Y_{l,m}(\theta_{k}, \phi_{k}) Y^{*}_{l',m'}(\theta_{k}, \phi_{k}) = \delta_{l,l'} \ \delta_{m,m'} \\ 
    &\sum_{l=0}^{\infty} \sum_{m=-l}^{l} Y_{l,m}(\theta_{1}, \phi_{1}) \ Y^{*}_{l,m}(\theta_{2}, \phi_{2}) = \frac{1}{\sin{\theta_{1}}} \ \delta(\theta_{1}- \theta_{2}) \delta(\phi_{1}-\phi_{2})
\end{align}
The integrand $\intk e^{i k \cdot x}$ becomes: 
\begin{align}\label{eq:integrand}
    \intk e^{i k \cdot x}= \sum_{l,m} 4 \pi i^{l} \bigg( \int_{0}^{\infty} d|k| \ |k|^{2} \ A_{lm}(|k|) \bessel \ \bigg) Y_{lm}(\theta_{x}, \phi_{x}).
\end{align}
The integrals which have to be performed explicitly consist in the computation of the angular coefficients in Eq. \ref{eq:angularintegration} and the integration in the radial compinent of $k$ in Eq. \ref{eq:integrand}. A change of variable is made in Eq. \ref{eq:integrand}: from $|k|$ to $u \equiv \beta$, through the relation $pc =(m c^{2}) \ u/(\hbar c)$, so that the integration can be performed over $u,\cos{\theta_{k}}, \phi_{k}$ in the domains: $[0,1],[0,1],[0,2\pi]$. The integration in $\phi$ is performed with the rectangle method, while the integration in $|k|, \cos{\theta_{k}}$ is performed with the Gaussian-Lagrange method, which for a generic function $f(x)$ states that \cite{gaussianq}: 
\begin{align}
    \int_{-1}^{1} f(x) dx = \sum_{i}^{N} \omega_{i} \ f(x_{i})
\end{align},
where the integral is approximated by a sum involving $2N$ parameters: $\{ \omega_{i} \}, \{ x_{i} \}$. The set $\{ x\}_{i=1}^{N}$ is composed by the roots of the Legendre polynomial of order $N$, namely $P_{N}(x)$. the weights $\{\omega_{i} \}_{i=1}^{N}$ are instead defined in the following way: 
\begin{equation}
    \omega_{i} = \int_{-1}^{1} \prod_{\substack{j=1 \\ j \neq 1}} \frac{x-x_{j}}{x_{i}-x_{j}} \ f(x_{i}).
\end{equation}

We define the DP electric field to which the EM eigenstates of the cavity couple in a generic coordinates system (${x,y,z}$):
\begin{equation}
\begin{split}
    E_{A'}(x) &= \chi m_{A}^{2} \partial_{t} A^{\mu} (x) \\
    A^{\mu} (x) &= \sum_{\lambda} \int \frac{d^{3} k }{(2 \pi)^{3} \sqrt{2 E(k)}} (\epsilon^{\mu}(\Vec{k}, \lambda) a(\Vec{k}, \lambda) e^{ikx} +  \epsilon^{\mu * }(\Vec{k}, \lambda) a^{\dagger}(\Vec{k}, \lambda) e^{-ikx})
\end{split}
\end{equation}
We now combine the two expressions and look at the non-relativistic limit of this field. The non-relativistic limit corresponds to the physical cases where the number operator $\mathcal{N}$ takes large numbers, so that the dispersion of the number operator $\Delta \mathcal{N}$  and the dispersion of the phase operator $\Delta \phi$ satisfy $\Delta \mathcal{N} \Delta \phi > \frac{1}{2}$ 
but the phase $\phi$ can be determined with high precision, given that $1 << \Delta \mathcal{N} << \mathcal{N}$. In this limit the creation and annihilation operator are complex number:
\begin{equation}
    a(\Vec{k}, \lambda) = |A(\Vec{k}, \lambda)| exp(i \theta(\Vec{k}, \lambda))
\end{equation}
where 
\begin{equation}
    \mathcal{N}(\Vec{k}, \lambda) = |A(\Vec{k}, \lambda)|^{2} 
\end{equation}
Giving an expectation value for the phase $\theta$ is a difficult conceptual task. $\theta$ will be the central value of the distribution of the spectrum of the operator $e^{i \phi}$ which is the unitary operator corresponding to anticylic permutation of the states with occupation number $n$. Given a value for the phase $\theta$ or equivalently for the space-distribution of the field $\phi$, it is always possible to define a probability distribution for the values in the spectrum of the operator $e^{i \phi}$. On the contrary, given an initial phase distribution, one should evaluate the classical limit of the permutation operator on the Fock states, which should eventually depend on the maximum occupation number. A relation for the half width of $\phi$ can be established when the NR field is a coherent state:
\begin{equation}
    \Delta \Phi \Delta N=\frac{1}{2} \rightarrow \Delta \Phi = \frac{1}{2 \sqrt{\Bar{N}}}
\end{equation}
Local DM density is $$
    \rho_{DM} = \frac{0.3 \text{GeV}}{cm^{3}}$$, so the number of dark non relativistic photons should be quite approximately $\sim \frac{\rho_{DM}}{m_{\gamma'}} V_{cavity}$. For a volume $\sim 1 \ m^{3}$ and a mass of $1 \ \mu eV$ we should have $10^{20} = \Bar{N}$ DPs. So we will assume that $\phi$ is distributed as a Dirac delta function, centered in an unknown central value $\phi_{0}$ which will be set to 0. \\
The creation and annihilation operators will be substituted: $$a(k), a^{\dagger}(k) \rightarrow \sqrt{\Bar{N (k)}} = \sqrt{\Bar{N P(k)}}$$. The DP field coupling to the EM field of the cavity is: 
\begin{equation}
    \Vec{E'}(x,t) = \partial_{t} \sum_{\lambda} \int \frac{d^{3} k }{(2 \pi)^{3} \sqrt{2 E(k)}} \sqrt{\Bar{N} P(k)}(\epsilon^{\mu}(\Vec{k}, \lambda)  e^{-ikx + i \omega t}+  \epsilon^{\mu * }(\Vec{k}, \lambda) e^{ikx - i \omega t} \big)
\end{equation}
The time domain of the field is controlled by the expression for $E=\sqrt{p^{2}+m^{2}_{\gamma'}}$. As for the time dependence, the expression is reduced to the 0 order term in the non relativistic expansion limit $m_{\gamma'} \gg k$: $$E(k) \sim m_{\gamma'} = E$$. The time dependence is independent from k and can be brought out of the integral, obtaining:
\begin{align}
    \Vec{E'}(x,t) &=  i  m_{\gamma'}^{1/2} \sum_{\lambda} \int \frac{d^{3} k }{(2 \pi)^{3} \sqrt{2} } \sqrt{\Bar{N} P(k)}\big(\epsilon^{\mu}(\Vec{k}, \lambda)  e^{-ikx + i m_{\gamma'} t} -  \epsilon^{\mu * }(\Vec{k}, \lambda) e^{ikx - i m_{\gamma'}t} \big)= \\ & i m_{\gamma'}^{1/2} \sum_{\lambda} \int \frac{d^{3} k }{(2 \pi)^{3} \sqrt{2} } \sqrt{\Bar{N} P(k)} Im(\epsilon^{\mu}(\Vec{k}, \lambda)  e^{-ikx + i m_{\gamma'} t}) =   \\ & i m_{\gamma'}^{1/2}\frac{\sqrt{\Bar{N}}}{(2 \pi)^{3} \sqrt{2}} \text{Im} \sum_{\lambda} \int d^{3} k \sqrt{P(k)} \epsilon^{\mu}(\Vec{k}, \lambda)  e^{-ikx + i m_{\gamma'} t}, 
\end{align}
where the operator $\sqrt{\Bar{N}(x)}$ can be rewritten as $\sqrt{\rho_{DM}(x) V}$. \\
Once the general expression for a non-relativistic photon field has been established, the probability $P(k)$ has to be defined. Asymptotically flat rotation cruves of the galaxies can be obtained with a Gaussian distribution of velocities (SHM) of the DM particles \cite{Evans:2018bqy}:
\begin{align}\label{eq:shm}
\fR(\vbf) = \frac{1}{(2\pi
  \sigma_v^2)^{3/2}N_\mathrm{R,esc}} \, \exp \left( - \frac{|\vbf|^2}{2\sigma_v^2}\right) \,
\nonumber\\ \times \Theta (\vesc - |\vbf|)\,,
\end{align}
\begin{equation}\label{eq:norm}
N_\mathrm{R,esc} = \erf \left( \frac{\vesc}{\sqrt{2}\sigma_v}\right) -
\sqrt{\frac{2}{\pi}} \frac{\vesc}{\sigma_v} \exp \left(
-\frac{\vesc^2}{2\sigma_v^2} \right)\,.
\end{equation}
where~$\sigma_v$ is the isotropic velocity dispersion of the DM and
$v_0 = \sqrt{2} \sigma_v$ is the value of the asymptotically flat rotation curve, while the constant $N_\mathrm{R,
esc}$ is used to renormalize the velocity distribution after
truncation to ~$\vesc$.

\bibliography{references}
\end{document}